\documentclass[pra, 12pt, showpacs, showkeys, eqsecnum]{revtex4}
\usepackage{graphics}

\newcommand {\drdtb}{\overline{d\rho /dt}}
\newcommand {\Tr} {{\mbox{Tr}}}

\begin{document}
\title{Assumptions that imply quantum dynamics is linear}
\author{Thomas F. Jordan}
\email[email: ]{tjordan@d.umn.edu}
\affiliation{Physics Department, University of Minnesota, Duluth, Minnesota 55812}

\begin{abstract}
A basic linearity of quantum dynamics, that density matrices are mapped linearly to density matrices, is proved very simply for a system that does not interact with anything else. It is assumed that at each time the physical quantities and states are described by the usual linear structures of quantum mechanics. Beyond that, the proof assumes only that the dynamics does not depend on anything outside the system but must allow the system to be described as part of a larger system. The basic linearity is linked with previously established results to complete a simple derivation of the linear Schrodinger equation. For this it is assumed that density matrices are mapped one-to-one onto density matrices. An alternative is to assume that pure states are mapped one-to-one onto pure states and that entropy does not decrease. 
\end{abstract}


\maketitle

\section{Introduction}\label{one}
Can we prove that quantum dynamics must be linear? That the Schrodinger equation must be linear? What assumptions are needed? Is there a reason in principle that quantum dynamics is linear? If not, quantum dynamics could be a linear approximation of a nonlinear theory; then experiments might reveal small nonlinear effects. If linearity can be proved from first principles, most physicists will not consider nonlinearity as an option for theories or spend time and money on experiments to look for nonlinear effects.

A definition of basic linearity for quantum dynamics is that density matrices are maped linearly to density matrices in an interval of time. This basic linearity is proved very simply here for a system that does not interact with anything else. It is assumed that at each time the physical quantities and states are described by the usual linear structures of quantum mechanics. What is proved is the basic linearity for the changes of states and mean values in time. Beyond the structure at each time, the proof assumes only that the dynamics does not depend on anything that happens outside the system, or on the state of any other system, but must allow the system to be described as part of a larger system. The latter condition is close in spirit to the idea of envariance\cite{Zurek05}, but the proof does not use quantum entanglement. The proof does not even require quantum mechanics; it could be done in classical mechanics as well.

This proof of basic linearity is the first of two steps. It is done in Section \ref{two}. The second step, taken in Section \ref{three}, links this basic linearity with previously established results to complete a simple derivation of the linear Schrodinger equation. For this it is assumed that the dynamics maps density matrices one-to-one onto density matrices in an interval of time. An alternative is to assume that pure states are mapped one-to-one onto pure states and that entropy does not decrease.

The question of the linearity of quantum dynamics has a rich history. Wigner proved\cite{WignerGroupTheory,Bargmann64} that quantum dynamics must be linear if it does not change absolute values of inner products of state vectors. Bialynicki-Birula and Mycielski\cite{BirulaMycielski} proposed a nonlinear Schrodinger equation that inspired precise experimental tests\cite{Shimony,ShullEtAl,GahlerEtAl}. Weinberg\cite{WeinbergNlqAnn,WeinbergNlqPrl} proposed a general nonlinear form of quantum mechanics, which led to more experimental tests\cite{Bollinger,Chupp,Walsworth,Majumder} and theoretical discussion\cite{PeresEntropy,WeinbergEntropy,me59,Polchinski,Czachor,me64,me66,GisinHPA,GisinExample,Svetlichny}. The seriousness of these proposals and their experimental tests shows that Wigner's proof of linearity was not conclusive; his assumption was questioned. I described where the proofs of linearity stood\cite{me60} at the time of Weinberg's proposal. Soon thereafter, an idea was proposed that relativity requires quantum dynamics to be linear\cite{GisinHPA,GisinExample,GisinX3}. This is discussed in Section \ref{four}.  

\section{Linear quantum dynamics}\label{two}

The result of quantum dynamics is the time dependence of mean values for Hermitian operators representing physical quantities. This includes the time dependence of probabilities, which are mean values for projection operators. The result is the same whether it is obtained from the Schrodinger picture or the Heisenberg picture. The basic linearity to be considered here is that the equations of motion for the mean values are linear in that the time derivative of each mean value is a function of mean values that depends on the state in a linear way. This does not mean that the equations of motion for operators are linear. The time derivative of an operator may be a product of operators, so the time derivative of the mean value of the operator is the mean value of a product of operators. The basic linearity to be proved here is just that the time derivative of a mean value does not involve products of mean values. It depends on the state in a linear way. This is the case in ordinary linear quantum mechanics. Proposals for nonlinear quantum mechanics provide examples with equations of motion where time derivatives of mean values are functions of mean values that depend on the state in a nonlinear way\cite{me64,me66}. They involve products of mean values. That happens because the Hamiltonian operator contains mean values; it depends on the state.

The mean value $\langle Q \rangle$ for a Hermitian operator $Q$ is $\Tr \left[ Q \rho \right]$ where $\rho $ is the density matrix that represents the state. In the Schrodinger picture, the time derivative of $\langle Q \rangle$ is
\begin{equation}
\label{eq:ld1}
\frac{d}{dt} \langle Q \rangle = \Tr \left[ Q \frac{d\rho} {dt} \right] .
\end{equation}
At each time, the equations of motion give time derivatives $d \langle Q \rangle /dt$ as functions of mean values at that time. The mean values at that time can be changed physically only by changing the state at that time. The basic linearity considered here is that each time derivative $d\langle Q\rangle /dt$ is a function of mean values that depends on the state in a linear way. The basic linearity fails if and only if there is a density matrix
\begin{equation}
\label{eq:ld2}
\rho = p\rho_{1} + (1-p)\rho_{2},
\end{equation}
 where $\rho_{1}$ and $\rho_{2}$ are density matrices and $p$ is a number between $0$ and $1$, such that $d\rho /dt$ is not the same as
\begin{equation}
\label{eq:ld3}
\overline{\frac{d\rho} {dt}} =p\frac{d\rho_{1}} {dt} + (1-p)\frac{d\rho_{2}} {dt}.
\end{equation}
The basic linearity is that $d\rho /dt$ and $\overline{d\rho /dt}$ are always the same. This means that density matrices are mapped linearly to density matrices in an interval of time. It is shown in Section \ref{three} that with additional assumptions this leads to the linear form of the Schrodinger equation.

Here is a proof that $d\rho /dt$ and $\overline{d\rho /dt}$ must always be the same. It assumes the dynamics is for a system $S$ that does not interact with anything else. Hence it assumes the dynamics does not depend on anything that happens outside $S$, or on the state of any other system. In particular, this means the dynamics can depend on the state of $S$ but not on any other property of the state of a larger system that contains $S$. It is also assumed that the dynamics must allow $S$ to be described as part of a larger system.

Suppose $S$ is one of two separate systems $S$ and $R$ and that the state of the larger system of $S$ and $R$ combined is represented by the density matrix
\begin{equation}
\label{eq:ld4}
\overline{\Pi} = p\rho_1|\alpha \rangle \langle \alpha | + (1-p)\rho_2|\beta \rangle \langle \beta |
\end{equation}
where $|\alpha \rangle$ and $|\beta \rangle$ are orthonormal vectors for $R$ that do not depend on the time and, as before, $\rho_1$ and $\rho_2$ are density matrices for $S$ and $p$ is a number between $0$ and $1$. The reduced density matrix $\Tr_R \overline{\Pi }$, which is the density matrix $\rho $ for $S$, is described by Eq.(\ref{eq:ld2} ). The probability $\langle P \rangle $ for a proposition represented by a projection operator $P$ for $S$ is the sum of joint probabilities
\begin{eqnarray}
\label{eq:ld5}
\langle P\rangle & = & \Tr_{SR}[P\overline{\Pi}] = \Tr_{SR}[P|\alpha \rangle \langle \alpha |\overline{\Pi}]  +  \Tr_{SR}[P|\beta \rangle \langle \beta |\overline{\Pi}] \nonumber \\ & = & \langle P|\alpha \rangle \langle \alpha |\rangle  +  \langle P|\beta \rangle \langle\beta |\rangle
 = p\Tr_S[P\rho_1 ]  +  (1-p)\Tr_S[P\rho_2]. 
\end{eqnarray}
Suppose a measurement is made on $R$ that distinguishes the states represented by $|\alpha \rangle$ and $|\beta \rangle$. The probability is $p$  that the result is $|\alpha \rangle$  and $1-p$ that the result is $|\beta \rangle$. If the result is $|\alpha \rangle$, the probability for the proposition represented by $P$ is $\Tr_S[P\rho_1 ]$, and if the result is $|\beta \rangle$, the probability for $P$ is $\Tr_S[P\rho_2 ]$. This can be verified experimentally by repeating the process of preparing the state represented by $\overline{\Pi}$, measuring to distinguish the states of $R$, and testing various propositions for $S$. The density matrices $\rho_1$ and $\rho_2$ describe physically distinct possibilities. The times of the events can be changed as long as the measurement on $R$ is early enough. The time dependence of $\rho_1$ and $\rho_2$ will account for changes in the results. The time derivative of the probability for the proposition represented by $P$ is $\Tr_S[Pd\rho_1 /dt]$ or $\Tr_S[Pd\rho_2 /dt]$ depending on the result of the measurement on $R$. Altogether, with the probabilities for both results being considered, the time derivative of the probability for $P$ is $\Tr_S[P\overline{d\rho /dt}]$.

The probability $\langle P\rangle$ for the proposition represented by $P$ is also
\begin{equation}
\label{eq:ld6}
\langle P\rangle =\Tr_{SR}[P\overline{\Pi}] = \Tr_S[P\Tr_R\overline{\Pi}] = \Tr_S[P\rho ].
\end{equation}
Its time derivative is $\Tr_S[Pd\rho /dt]$. This is always correct. It may be the only possibility at hand. For example, suppose the state of $S$ and $R$ combined is represented by the density matrix
\begin{equation}
\label{eq:ld7}
\Pi = p\rho |\alpha \rangle \langle \alpha | + (1-p)\rho |\beta \rangle \langle \beta | = \rho [p|\alpha \rangle \langle \alpha | + (1-p)|\beta \rangle \langle \beta |].
\end{equation}
The dynamics for $S$ must be the same for $\Pi$ as for $\overline{\Pi}$.

The time derivative of the probability for the proposition represented by $P$ is always $\Tr_S[Pd\rho /dt]$. There are situations where it also must be $\Tr_S[P\overline{d\rho /dt}]$ to fit observations of events in a larger system. The dynamics in $S$ can not depend on the situation of $S$ in a larger system. Therefore $\Tr_S[Pd\rho /dt]$ and $\Tr_S[P\overline{d\rho /dt}]$ must be the same. From that equality for various projection operators $P$, we conclude that $d\rho /dt$ and $\overline{d\rho /dt}$ must be the same.

Nothing in this proof requires quantum mechanics. Quantum entanglement is not used. The correlations involved are classical. The substance would be the same in classical mechanics; only the language would be different.

\section{The Schrodinger equation}\label{three}

The result of Section \ref{two} brings us within sight of a proof that the dynamics can be described by changes of state vectors in time determined by a linear Schrodinger equation. The classic proof assumes that in an interval of time the dynamics maps the set of all pure states one-to-one onto itself. The first step in the proof is Wigner's theorem that if the absolute values of inner products of state vectors do not change, the change of state vectors can be made with an operator that is either linear or antilinear\cite{WignerGroupTheory,Bargmann64}. The product of two antilinear operators is linear, so if the change of state vectors from one time to another can be made in two steps, it must be linear\cite{meLinearOperatorsChapter6}. If the change of states does not depend on the time when the change begins, and if the change of probabilities in time is continuous, then the state vectors as functions of time can be obtained with a continuous one-parameter group of unitary operators\cite{wigner39a,Bargmann54,meLinearOperatorsChapter6}. They satisfy a linear Schrodinger equation; the Hermitian generator is the linear Hamiltonian operator\cite{meLinearOperatorsChapter6}.

If the dynamics applies to all states, it must map every density matrix $\rho $ to a density matrix $\rho '$ in an interval of time. If density matrices $\rho_1$ and $\rho_2$ are mapped to $\rho_1'$ and $\rho _2'$, then what was proved in Section \ref{two} implies that the density matrix described by Eq.(\ref{eq:ld2} ) is mapped to
\begin{equation}
\label{eq:se1}
\rho ' = p\rho_1' + (1-p)\rho_2'.
\end{equation}
The map is linear for density matrices. It has been known for some time that this can be used to prove the assumptions of Wigner's theorem and open the way to the linear form of the Schrodinger equation\cite{jordan62,Kadison,Hunziker,Simon}.

Linear maps of density matrices can also be used to describe processes where pure states are mapped to mixed states, different states are mapped to the same state, the map is not onto all states or, generally, the map has no inverse that applies to all states. An assumption is needed to separate these processes from dynamics described by the Schrodinger equation, which has an inverse for all states.

Suppose that in an interval of time the dynamics maps the set of all density matrices one-to-one onto itself. The result of Section \ref{two} is assumed, that the map is linear for density matrices, as described by Eq.(\ref{eq:se1}) for a density matrix described by Eq.(\ref{eq:ld2}). Then the map has an inverse and the inverse map is linear; the proof that the inverse of a linear operator is linear\cite[Theorem 7.1]{meLinearOperatorsChapter6} applies with attention restricted to density matrices. Pure states are mapped to pure states: if $\rho_1'$ and $\rho_2'$ in Eq.(\ref{eq:se1}) are distinct, so are $\rho _1$ and $\rho _2$ in Eq.(\ref{eq:ld2}); thus if $\rho '$ is for a mixed state, so is $\rho $. The inverse map also takes pure states to pure states, so the set of all pure states is mapped one-to-one onto itself.

For each vector $|\psi \rangle$ of length $1$, let $|\psi '\rangle$ be a vector of length $1$ such that $(|\psi \rangle \langle \psi |)'$ is $|\psi '\rangle \langle\psi '|$. For each density matrix $\rho $ there are orthonormal vectors $|\psi_j \rangle$ and positive numbers $p_j$ whose sum $\sum_j p_j$ is $1$ such that
\begin{equation}
\label{eq:se2}
\rho  = \sum_j p_j|\psi_j \rangle \langle \psi_j |.
\end{equation}
The linearity implies that
\begin{equation}
\label{eq:se3}
\rho ' = \sum_j p_j |\psi_j' \rangle \langle \psi_j' |.
\end{equation}
Since $\langle \psi_j' |\psi_j' \rangle$ is $1$,
\begin{equation}
\label{eq:se4}
\Tr[(\rho' )^2] = \sum_{jk} p_j p_k |\langle \psi_j' |\psi_k' \rangle |^2 \geq \sum_j (p_j)^2 = \Tr[\rho^2 ].
\end{equation}
The same result for the inverse implies that
\begin{equation}
\label{eq:se5}
\Tr[(\rho ')^2] = \Tr[\rho^2].
\end{equation}
Let $|\psi \rangle$ and $|\phi \rangle$ be vectors of length $1$ and let
\begin{equation}
\label{eq:se6}
\rho  = \frac{1}{2} |\psi \rangle \langle \psi | + \frac{1}{2} |\phi \rangle \langle \phi |.
\end{equation}
Then
\begin{equation}
\label{eq:se7}
Tr[\rho ^2] = \frac{1}{2} + \frac{1}{2} |\langle \psi |\phi \rangle |^2
\end{equation}
and
\begin{equation}
\label{eq:se8}
\rho ' = \frac{1}{2} |\psi '\rangle \langle \psi '| + \frac{1}{2} |\phi '\rangle \langle \phi '|
\end{equation}
so Eq.(\ref{eq:se5} ) implies that
\begin{equation}
\label{eq:se9}
|\langle \psi '|\phi '\rangle |^2 = |\langle \psi |\phi \rangle |^2.
\end{equation}
Absolute values of inner products of state vectors are not changed. The assumptions of Wigner's theorem have been proved. The derivation of the linear form of the Schrodinger equation can proceed along the classic route.

An alternative is to assume that it is the set of all pure states that is mapped one-to-one onto itself, instead of the set of all density matrices. The result of Section \ref{two} is assumed as before, so the dynamics defines a map of density matrices to density matrices that is linear for density matrices. Then the inequality (\ref{eq:se4}) is obtained as before. If it is assumed that the entropy does not decrease\cite{PeresEntropy}, then
\begin{equation}
\label{eq:se10}
\Tr[(\rho ')^2] \leq  \Tr[\rho ^2].
\end{equation}
This implies Eqs.(\ref{eq:se5}) and (\ref{eq:se9}). An alternative to assuming that entropy does not decrease is to assume that the probabilities in mixtures of orthogonal pure states do not change\cite{me60}.  
 
\section{Discussion} \label{four}

Questions about previous proofs helped motivate this work. The idea that relativity might require quantum dynamics to be linear was clearly expressed in an example\cite{GisinExample} where nonlinear dynamics makes a $\drdtb$ defined by Eq.(\ref{eq:ld3}) different from $d\rho /dt$ and lets it be changed from outside the system. The system $S$ is in an entangled state with a separate system $R$ so that one of two possible results of a particular measurement in $R$ implies that probabilities of propositions for $S$ are described by the density matrix $\rho _1$, and the other possible result from $R$ implies that probabilities for $S$ are described by the density matrix $\rho_2$. Let $p$ and $1-p$ be the probabilities of the two possible results from $R$. The $\drdtb$ of Eq.(\ref{eq:ld3}) is not the same as $d\rho /dt$. A different measurement in $R$ yields different density matrices $\rho_1$ and $\rho_2$ and a different $\drdtb$. If $\drdtb$ describes the time dependence of probabilities for $S$, then observations in $S$ can determine which measurement was made in $R$. A signal can be sent from $R$ to $S$. Since there is no restriction on the locations of $R$ and $S$, the signal can be faster than light. This assumes that $\drdtb$, not $d\rho /dt$, describes the time dependence of probabilities for $S$. This assumption can be questioned\cite{me64,me71}.

If a signal can not travel from $R$ to $S$ at the speed of light in the time between the measurement in $R$ and observations in $S$, then moving observers can disagree about whether the observations in $S$ happen after or before the measurement in $R$. The probabilities for the results of the measurement in $R$ are calculated from the state of the larger system of $S$ and $R$ combined, without consideration of what happens in $S$. Probabilities for $S$ can be calculated the same way, from the density matrix $\rho $ for $S$ that is the reduced density matrix obtained by taking the trace for $R$ of the density matrix for $S$ and $R$ combined, and is related to $\rho_1$, $\rho_2$, $p$ and $1-p$ by Eq.(\ref{eq:ld2}). The state of $S$ and the state of $R$ both are prepared when the state of $S$ and $R$ combined is prepared. Can we not just say\cite{me64,me71,CzachorDoebner} that the time dependence of probabilities for $S$ is described by $d\rho /dt$, not $\drdtb$, in this situation? Then there would be no signal faster than light. Use of $\drdtb$ could be reserved for mixtures where $d\rho_1/dt$ and $d\rho_2/dt$ have to be calculated separately because the states represented by $\rho_1$ and $\rho_2$ are definitely prepared separately\cite{me64}. Should we not be prepared to refine our rules of interpretation of quantum mechanics just this way if experiments show evidence of nonlinear quantum dynamics?

If there is time for a signal to travel from $R$ to $S$ at the speed of light and report the result of the measurement in $R$ before the observations are made in $S$, we can say that a state of $S$ represented by $\rho_1$ or $\rho_2$ is definitely prepared before the observations, so that altogether, when the probabilities $p$ and $1-p$ for both possibilities are considered, the time dependence of probabilities for $S$ should be described by $\drdtb$. All of this would be simpler in the context of nonrelativistic mechanics where moving observers agree on the order of events in time and there is no limit on the speed of signals. Could relativity actually provide a niche where nonlinear quantum dynamics can exist? If there were nonlinear quantum dynamics, there would have to be an abrupt change from $\drdtb$ to $d\rho /dt$ when gradual changes in distances and times reach the point where a signal can no longer travel from $R$ to $S$ at the speed of light in the time between the measurement and observations\cite{me71}. This makes nonlinear quantum dynamics look unreasonable, largely unattractive for investment of time. Does it prove that nonlinear quantum dynamics is impossible? That it does not can be demonstrated with a fanciful construction.\cite{Kent}

This example uses Weinberg's nonlinear quantum mechanics. When the idea that relativity implies linearity was presented more generally\cite{GisinHPA, GisinX3}, the proofs assumed that if a map of density matrices to density matrices is defined by the dynamics, it must be linear in mixtures of pure states. This does not hold for Weinberg's nonlinear quantum mechanics\cite{WeinbergNlqAnn,WeinbergNlqPrl} formulated in terms of density matrices\cite{me64,me66}, so these proofs do not effectively address Weinberg's proposal; it is simply ruled out by that assumption at the start. The same assumption was made to present the idea that nondecreasing entropy requires linearity\cite{PeresEntropy}. Using entropy together with basic linearity as in Section \ref{three} simplifies the proof using entropy and makes it clear that the only additional assumption that is needed\cite{me60} is that pure states are mapped one-to-one onto pure states.

The first presentation of the idea that relativity requires linearity\cite{GisinHPA} stops when a linear map of density matrices is obtained. The more recent presentation\cite{GisinX3} goes on to a proof that the map of density matrices is completely positive, so that it ``can be realized ... by a linear and unitary evolution on a larger Hilbert space.'' This would include dynamics that is not described by a Schrodinger equation but is for a subsystem of a larger system where the dynamics is described by a Schrodinger equation. It would bring in cases where the considered system interacts with other parts of the larger system. We know now that evolution in a subsystem caused by linear unitary evolution in a larger system is described by linear maps of density matrices that generally are not completely positive and act in limited domains\cite{jordan04}. Complete positivity generally will not provide a link to the source of the dynamics. Including subsystem dynamics will be complicated by the need to consider domains. It is simpler to consider only the given system and make the assumptions needed to link up with Wigner's theorem as in Section \ref{three}.

\section*{ACKNOWLEDGMENT}

I am grateful to Wojciech Zurek for a suggestion that brought me back to this subject with my eyes open.

\bibliography{ncp}
\end{document}